\documentclass[12pt,a4paper]{article}
\usepackage{amsmath}
\usepackage{amssymb}
\usepackage{enumerate}
\usepackage{amsfonts}
\usepackage[latin1]{inputenc}
\begin{document}

%
\newcounter{saveeqn}
\newcommand{\alpheqn}{\setcounter{saveeqn}{\value{equation}}%
\setcounter{equation}{0}%
\addtocounter{saveeqn}{1}
\renewcommand{\theequation}{\mbox{\arabic{saveeqn}\alph{equation}}}}
\newcommand{\reseteqn}{\setcounter{equation}{\value{saveeqn}}%
\renewcommand{\theequation}{\arabic{equation}}}
\begin{center}
{\bf  ``Light-Front Quantization of the Nielsen-Olesen (Bogomol'nyi) Model'' \footnote{ ``Invited Contributed Talk'' at `` Workshop on Light-Cone 2007 Relativistic Hadronic and Nuclear Physics'', Ohio Center for Technology and Science, Ohio State University, Columbus, Ohio, USA, May 14-18, 2007.}   }
\\[14mm]

Usha Kulshreshtha$^{[{\rm a}]}$ 

\end{center}

\begin{tabular}{cc}
a & Department of Physics, Kirori Mal College, University of Delhi,\\
& Delhi-110007, India. Email: $<$ushakulsh@gmail.com$>$ \\
\end{tabular} 

\vspace{0.5cm}

\begin{abstract}
In this work, I consider the light-front quantization of a class of Nielsen-Olesen (Bogomol'nyi) models in two-space one-time dimensions in the so-called symmetry phase using the Hamiltonian, path integral and BRST formulations.
 
\end{abstract}

In this work, I study a class of Nielsen-Olesen (Bogomol'nyi) models\cite{1,2} in two-space one-time dimensions in the so-called symmetry phase using the Hamiltonian, path integral and BRST formulations\cite{3} on the light-front (LF) (i.e., on the the hyperplanes of the LF with the light-cone (LC) time ( $ x^{+}\equiv \tau ) = ( x^{0} + x^{1})/ \sqrt{2} = $ constant ) describing the FF dynamics\cite{4} which is in different than the instant-form (IF) dynamics of the system on the on the hyperplanes: $ x^{0}$ = t = constant)\cite{4}. These models have a  Maxwell term in the action which accounts for the Kinetic energy of the vector gauge field A$^{\mu}$ (x$^{\mu}$), and they represent field theoretical models which could be considered as effective theories of the Ginsburg-Landau-type for superconductivity\cite{1,2} are, in fact, the relativistic generalizations of the well known Ginsburg-Landau phenomenological field theory models of superconductivity \cite{2} and are known as the Nielsen-Olesen (vortex) models (NOM)\cite{1,2}. This class of models is defined by the following action \cite{1,2}:
\alpheqn
\begin{eqnarray}
S &=& \int_{ }^{} {\cal L}(\Phi , \Phi^{\ast}, A^{\mu}) d^{3}x ;\quad
{\cal L} = \biggl[ -\frac{1}{4} F_{\mu\nu} F^{\mu\nu} +
(\tilde{D}_{\mu}\Phi^{\ast})(D^{\mu}\Phi) -V(|\Phi|^{2})\biggr] \\
\label{1a}
V(|\Phi|^{ 2}) &=& [\alpha _{0 }+ \alpha_{ 2}|\Phi|^{ 2}+ \alpha_{4}|\Phi|^{4} ] \quad = [ \lambda (|\Phi|^{2}- \Phi^{2}_{0})^{2}  ] \quad;\quad 
\Phi_{0}\ne 0    \\
\label{1b}
F_{\mu\nu} &=& (\partial_{\mu} A_{\nu }- \partial_{\nu} A_{\mu});\quad
D_{\mu} = (\partial_{\mu} + ieA_{\mu});\quad \tilde{D}_{\mu} =
(\partial_{\mu} - ieA_{\mu}) \\
\label{1c}
{\rm g}^{\mu\nu} & := & {\rm diag}(+1, -1, -1)\quad ; \quad \mu,\nu = 0,1,2.    \label{1d}
\end{eqnarray}
\reseteqn
If the parameters of the Higgs potential are chosen such that the masses of scalar (Higgs boson) and that of the  vector spin one particle (photon) become equal, i.e., if one has:\cite{1,2}:
\begin{equation}
m_{Higgs}  =  m_{photon}    = e  \Phi_{0} \quad;\quad \lambda = \frac{1}{2} e^{2} \quad;\quad 
V(|\Phi|^{ 2}) = \frac{1}{2} e^{2}  (|\Phi|^{2} - \Phi^{2}_{0})^{2} 
\label{2}
\end{equation}
then above NOM reduces to the  so-called  Bomol'nyi  model\cite{2,3} which  describes a system on the boundary between type-I and type-II superconductivity and admits self dual solitons\cite{1,2}. The parameters of the Higgs potential are choosen such that the potential remains a double well potential with $\Phi_{ 0}\ne$ 0. In this case the spontaneous symmetry breaking takes place due to the non-invariance of the lowest (ground) state of the system (because  $\Phi_{0} \ne 0  $ ) under the operation of the local $ U(1) $ symmetry, and the symmetry which is broken is still a symmetry of the system and it is manifested in a manner other than the invariance of the lowest Ground) state ($\Phi_{0} $) of the system. However, no Goldstone boson occures here and instead the gauge field acquires a mass through some kind of a Higgs mechanism and the symmetry is manifested in the Higgs mode. The model is easily seen to  possess five constraints:
\alpheqn
\begin{eqnarray}
\chi_{ 1} &=& \Pi^{ +} \approx 0 \quad ; \quad 
\chi_{2}  =  [\Pi - \partial _{-}\Phi^{\ast} + ieA^{+}\Phi^{\ast} ]\approx 0  \\\label{3a}
\chi_{3} &=& [\Pi^{\ast} - \partial _{-}\Phi - ieA^{+}\Phi ]\approx 0 
\quad ;\quad
\chi_{4}= [E + \partial_{2}A^{+} - \partial_{-}A_{2}] \approx 0  \\   
\label{3b}
\chi_{ 5} &=& [ ie (\Pi\Phi - \Pi^{\ast} \Phi^{\ast}) + \partial_{2} E + \partial_{-}\Pi^{-} ] \approx 0
\label{3c}
\end{eqnarray}
\reseteqn
where  $\Pi$ , $\Pi^{\ast}$, $\Pi^{+}$, $\Pi^{ -}$ and $ E(:= \Pi^{ 2})$  are the momenta canonically conjugate respectively to $\Phi$, $\Phi^{\ast}$,  A$^{-}$,A$^{+}$ and $ A_{2}$. Here the first four constraints are primary and the last one is the secondary Gauss law constraint. The total Hamiltonian density ${\cal H}_{T}$  of the theory is obtained by including the primary constraints in the canonical Hamiltonian density  with the help of the Lagrange multiplier fields $u$, $v$ and $w$. It is now easily seen that the constraints  $\chi_{2}, \chi_{3}, \chi_{4}$ and $\chi_{5}$ could be combined in to a single constraint: 
\begin{equation}
\eta = [ ie (\Pi\Phi - \Pi^{\ast} \Phi^{\ast}) + \partial_{2} E + \partial_{-}\Pi^{-} ] \approx 0
\label{4}
\end{equation} 
and with this modification, the theory is seen to possess a set of only two constraints: $ \Omega_{1} ( = \chi_{ 1})$ and $ \Omega_{2} ( = \eta ) $. Further, the matrix of the Poisson brackets among the constraints $\Omega_{1}$ and  $\Omega_{2}$ is seen to be a null matrix implying that the set of constraints $\Omega_{1}$ and   $\Omega_{2}$ is first-class and that the theory under consideration  is  gauge-invariant.  The action of the theory  is indeed seen to have a  local vector gauge symmetry and the divergence of vector gauge current density of the theory is easily seen to vanish (with $ \partial_{\mu} j^{\mu} $ = $  0  $).  The theory could be quantized e.g., under the set of light-cone gauges: $ A^{+} = 0 $ and $ A^{-} = 0 $. Finally, following the standard Dirac quantization procedure, the nonvanishing equal light-cone-time (ELCT) commutators of the theory, under these light-cone gauges could be easily obtained and are omitted here for the sake of brevity. In the path integral formulation, the transition to quantum theory is made  by writing the vacuum to vacuum transition amplitude for the theory  called the generating functional  $ Z[J_{k}] $ of the theory \cite {2,3} under the light-cone gauges under consideration, in the presence of the external sources $J_{k}$ \cite{3}:
\begin{eqnarray}
Z[J_{k}] &=& \int [d\mu] \exp \biggl [i \int d^{3} x \biggl [ J_{k} \Phi^{k} + \Pi \partial_{+}\Phi + \Pi^{\ast} \partial_{+}\Phi^{\ast}  + \Pi^{+}\partial _{+}A^{-} + \Pi^{-}\partial_{+}A ^{+}   \nonumber \\
&&  \quad  + E \partial_{+}A_{2}  + \Pi_{u}\partial_{+}u  + \Pi_{v}\partial_{+}v + \Pi_{w}\partial_{+} w  + \Pi_{z}\partial_{+} z - {\cal H}_{T} \biggr ] \biggr ] 
\label{5}
\end{eqnarray}
where the phase space variables of the theory are: $\Phi^{k} \equiv (\Phi, \Phi^{\ast}, A^{-}, A^{+}, A_{2}, u, v, w, z)$ with the corresponding respective canonical conjugate momenta: $\Pi_{k} \equiv (\Pi, \Pi^{\ast}, \Pi^{+}, \Pi^{-}, E , \Pi_{u},$ \\ $\Pi_{v},\Pi_{w}, \Pi_{z} ) $. The functional measure $[d\mu]$ of the generating functional $Z[ J_{k}]$ under the above LC gauges is obtained as\cite{3}: 
\begin{eqnarray}
[d\mu] &=&  [  [\partial_{-}\delta(x^{-} - y^{-})] [\delta(x^{-} - y^{-})][\delta^{2}(x_{2}-y_{2})] ] [d\Phi] [d {\Phi}^{\ast}]  [dA^{+}] [dA^{-}]  [dA_{2}] 
  [du]  [dv] \nonumber \\                
& & \qquad [dw ] [dz] [d\Pi]  [d\Pi{\ast}]  [d\Pi^{-}]  [d\Pi^{+}]  [dE]  [d\Pi_{u}] [d\Pi_{v}] [d\Pi_{w}]  [d\Pi_{z}]  \delta [ \Pi^{+} \approx 0] \nonumber \\& & \qquad \delta [ ( ie (\Pi\Phi - {\Pi}^{\ast}{\Phi}^{\ast}) + \partial_{2} E + \partial_{-}\Pi^{-} )\approx 0 ] \delta [A^{+} \approx 0] \delta [ A^{-} \approx 0] 
\label{6}
\end{eqnarray}
For the BRST formulation of the model, we first enlarge the Hilbert space of our gauge-invariant theory and replace the notion of gauge-transformation, which shifts operators by c-number functions, by a BRST transformation, which mixes operators with Bose and Fermi statistics, we then introduce new anti-commuting variable c and $ \bar{c} $ (Grassman numbers on the classical level, operators in the quantized theory) and a commuting variable $b$ such that\cite{4}:
\alpheqn
\begin{eqnarray}
\hat{\delta}\Phi &=& i c \Phi\quad , \quad \hat{\delta}\Phi^{\ast} = - i c \Phi^{\ast}\quad ,\quad \hat{\delta} A^{-} = - \partial_{+} c \quad ,\quad \hat{\delta} A_{2} = - \partial_{2} c   \\
\hat{\delta} A^{+} &=& - \partial_{-} c \quad ,\quad  
\hat{\delta} \Pi^{+}  =   \hat{\delta} \Pi^{-}  =  \hat{\delta} E  =  0 
\quad ,\quad \hat{\delta} z = - \partial_{+} \partial_{2} c   \\
\hat{\delta} \Pi &=& ( -i c \partial_{-}\Phi^{\ast} - e c A^{+} \Phi^{\ast} )  , \quad 
\hat{\delta} \Pi^{\ast} = ( i c \partial_{-}\Phi - e c A^{+} \Phi )\\
\hat{\delta} u &=& - \partial_{+} \partial_{+} c   \quad ,\quad 
\hat{\delta} v =  i (\Phi \partial_{+} c +  c \partial_{+} \Phi ),\quad  \hat{\delta} w =  - i (\Phi^{\ast} \partial_{+} c  +  c \partial_{+} \Phi^{\ast} )\\
\hat{\delta}\Pi_{u} &=& \hat{\delta}\Pi_{v} =  \hat{\delta}\Pi_{w} =  \hat{\delta}\Pi_{z} = 0 \quad ,\quad \hat{\delta}c = 0 \quad ,\quad \hat{\delta}\bar{c} = b \quad, \quad \hat{\delta}b = 0 
\label{7}
\end{eqnarray}
\reseteqn
with the property $\hat{\delta}^{2}$ = 0.  We now define a BRST-invariant function of the dynamical variables to be a function $ f $ such that $\hat{\delta} f =  0 $.  Now the BRST gauge-fixed quantum Lagrangian density ${\cal L}_{BRST}$ for the theory could be obtained by adding to the first-order Lagrangian density ${\cal L}_{I0}$, a trivial BRST-invariant function\cite{2,3}, e.g., as follows: 
\begin{eqnarray}
{\cal L}_{BRST} &:=& \biggl [ \Pi^{-} \partial_{+}A ^{+} - \Pi^{-} \partial_{-}A ^{-}  - \frac{e^{2}}{2} (\Pi^{-})^{2}  +
(\partial_{2}A^{+} - \partial_{-}A_{2}) (\partial_{2}A^{-} - \partial_{+} A_{2}) \nonumber \\ 
&& \qquad (\partial_{-}\Phi + i e A^{+}\Phi)\partial_{+}\Phi^{\ast}
 +  (\partial_{-}\Phi^{\ast} - i e A^{+}\Phi^{\ast})\partial_{+}\Phi
\nonumber \\
&& \qquad + ie\Phi A^{-}\partial_{-}\Phi^{\ast} - ie  A^{-} \Phi^{\ast} \partial_{-}\Phi + 2e^{2}A^{-} A^{+}\Phi^{\ast} \Phi  \nonumber \\      
&& \qquad - (\partial_{2} \Phi^{\ast})\partial_{2} \Phi - i e A_{2}\Phi
\partial_{2} \Phi^{\ast} + ie A_{2} \Phi^{\ast}\partial_{2} \Phi - e^{2} A_{2}^{2} \Phi^{\ast} \Phi \nonumber \\
&& \qquad  - V(|\Phi|^{2}) + \hat{\delta}[ \bar{c}(-\partial_{ +}A^{-}  + \frac{1}{2}b) ]  \biggr ]
\label{8}
\end{eqnarray}
The last term in the above equation is the extra BRST-invariant gauge-fixing term.  After one integration by parts, the above equation could now be written as:
Proceeding classically, the Euler-Lagrange equation for $b$ reads:
$  b =(\partial_{ +}A ^{-} ) $ and the requirement $\hat{\delta} b = 0 $ then implies $  \hat{\delta}b =[\hat{\delta}(\partial_{ +}A^{-}) ] $ which in turn implies $\partial_{ +}\partial_{ +}c = 0 $. We now define the bosonic momenta as: $ \Pi^{ +}:=  - b $  and  the fermionic momenta with directional derivatives as:$ \Pi_{ c}:= \partial _{+}\bar{c} $  and $ \Pi_{\bar{c}} := \partial_{+}c $, implying that the variable canonically conjugate to $c$ is ( $\partial_{+} \bar{c}$) and the variable conjugate to $\bar{c}$ is ($\partial_{+} c$).  The quantum Hamilotonian density $ {\cal H}_{BRST} $ of the theory could now be easily constructed and the corresponding Hamilton's equations for the bosonic and fermionic variables could be obtained. Also for the operators $c, \bar{c}, \partial_{+}c$ and $\partial_{+} \bar{c}$, one needs to satisfy the anticommutation relations of $\partial_{+} c$ with $\bar{c}$ or of $\partial_{+} \bar{c}$ with $c$, but not of $c$, with $\bar{c}$. In general, $c$ and $\bar{c}$ are independent canonical variables and are seen to possess the following properties:

\begin{eqnarray}
c^{2}  = {\bar{c}}^{2} = \{\bar{c},c\} = \{ \partial_{+}\bar{c}, \partial_{+}c\} = \{\Pi_{ c}, \Pi_{\bar{c}}\} = 0 \quad;\quad \{\partial_{+}\bar{c},c \} = (-1)\{\partial_{+}c,\bar{c}\} = i 
\label{9}
\end{eqnarray}

where $\{\makebox{ }$,  $\}$ means an anticommutator. The BRST charge operator $Q$ is the generator of the BRST transformations. It is nilpotent and satisfies $Q^{2}$ = 0. It mixes operators which satisfy Bose and fermi statistics. According to its conventional definition, its commutators with Bose operators and its anti-commutators with Fermi operators could be obtained rather easily and the BRST charge operator of the present  theory can be written as
\begin{equation}
Q = \int_{ }^{}\!dx^{-}dx_{2}
\biggl [ ic (\partial_{-}\Pi^{-} + \partial_{2} E + ie \Pi \Phi
- ie \Pi^{\ast} \Phi^{\ast} )  - i(\partial_{ +}c) \Pi^{+} \biggr ]
\label{10}
\end{equation}
This equation implies that the set of states satisfying the conditions: 
\begin{equation}
\Pi^{0}|\psi \rangle = 0 \quad,\quad \biggl [ \partial_{-} \Pi^{-} + \partial_{2} E + i e (\Pi\Phi - \Pi^{\ast}\Phi^{\ast}) \biggr ] |\psi \rangle = 0 
\label{11}
\end{equation}
belong to the dynamically stable subspace of states $|\psi>$ satisfying $Q|\psi> = 0 $, i.e., it belongs to the set of BRST-invariant states. The Hamiltonian is also invariant under the anti-BRST transformations (which are omitted here for the sake of brevity) with the generator or anti-BRST charge:
\begin{equation}
\bar{Q} = \int_{ }^{}\!dx^{-}dx_{2}
\biggl [ - i \bar{c} (\partial_{-}\Pi^{-} + \partial_{2} E + ie \Pi \Phi
- ie \Pi^{\ast} \Phi^{\ast} )  + i(\partial_{+}\bar{c}) \Pi^{+} \biggr ]
\label{12}
\end{equation}
We also have $ \partial_{ +}Q= [Q, H_{BRST}]= 0 $ and $ \partial_{ +}\bar{Q}=
[\bar{Q}, H_{BRST}]= 0  $ with $ H_{BRST}= \int_{ }^{} dx^{-} dx_{2} {\cal H}_{BRST} $ and we further impose the dual condition that both $Q$ and $\bar{Q}$ annihilate physical states, implying that $ Q |\psi > =0 $ and $ \bar{Q}|\psi> = 0 $. Now because $Q|\psi > = 0$, the set of states annihilated by $Q$ contains not only the set of states for which the constraints of the theory hold but also additional states for which the constraints of the theory do not hold in particular. This situation is, however, easily avoided by aditionally imposing on the theory, the dual condition:  $ \bar{Q}|\psi> = 0 $ and $  $ $ \bar{Q}|\psi> = 0  $. Thus by imposing both of these conditions  on the theory simultaneously, one finds that the states for which the constraints of the theory hold satisfy both of these conditions and, in fact, these are the only states satisfying  both of these conditions  because in view of the conditions on the fermionic variables $c $ and $\bar{c}$ one cannot have simultaneously $c$, $\partial_{+} c $ and $\bar{c}$, $\partial_{+}\bar{c}$, applied to $|\psi >  $ to give zero. Thus the only states satisfying  $ Q |\psi> = 0 $ and $  $ $ \bar{Q}|\psi> = 0  $  are those that satisfy the constraints of the theory and they belong to the set of BRST-invariant as well as to the set of anti-BRST-invariant states. Towards the end I would review the work of Refs.\cite{2} and compare it with the present results.

\end{document}